# An Investigation of Commercial Iron Oxide Nanoparticles: Advanced Structural and Magnetic Properties Characterization

Kai Wu[†,⊥,*], Jinming Liu[†,⊥], Renata Saha[†,⊥], Chaoyi Peng[†], Diqing Su[‡], Andrew Yongqiang Wang[§,*], and Jian-Ping Wang[†,*]

[†]Department of Electrical and Computer Engineering, University of Minnesota, Minneapolis, Minnesota 55455, USA

[‡]Department of Chemical Engineering and Material Science, University of Minnesota, Minneapolis, Minnesota 55455, USA

[§]Ocean NanoTech, LLC, San Diego, California 92126, USA

**ABSTRACT:** Magnetic nanoparticles (MNPs) have been extensively used as tiny heating sources in magnetic hyperthermia therapy, contrast agents in magnetic resonance imaging (MRI), tracers in magnetic particle imaging (MPI), carriers for drug/gene delivery, etc. There have emerged many magnetic nanoparticle/microbeads suppliers since the last decade, such as Ocean NanoTech, Nanoprobes, US Research Nanomaterials, Miltenyi Biotec, micromod Partikeltechnologie GmbH, and nanoComposix, etc. In this paper, we report the physical and magnetic characterizations on iron oxide nanoparticle products from Ocean NanoTech. Standard characterization tools such as Vibrating-Sample Magnetometer (VSM), X-Ray Diffraction (XRD), Dynamic Light Scattering (DLS), Transmission Electron Microscopy (TEM), and Zeta Potential Analyzer are used to provide magnetic nanoparticle customers and researchers with an overview of these iron oxide nanoparticle products. In addition, the dynamic magnetic responses of these iron oxide nanoparticles in aqueous solutions are investigated under low and high frequency alternating magnetic fields, giving a standardized operating procedure for characterizing the MNPs from Ocean NanoTech, thereby yielding the best of magnetic nanoparticles for different applications.

## 1. INTRODUCTION

Magnetic nanoparticles (MNPs) are nanomaterials with sizes between 1 nm and 100 nm. Due to their large surface-to-volume ratio and tunable magnetic properties, MNPs have emerged as one of the most important nanomaterials in magnetic, chemical and biomedical applications. The surface of the MNPs can be functionalized with various coatings from inorganic coatings such as silica[1] and carbon[2] to organic coatings such as polyethylene glycol (PEG)[3] and dopamine (DPA)[4]. Compared to non-magnetic particles, MNPs can be manipulated by an external magnetic field without any physical contact, which leads to various applications such as drug delivery[5] as well as the separation and concentration of certain molecules[6]. Under an alternating magnetic field, MNPs can induce localized temperature increase at the target spot, which makes them promising candidates for hyperthermia applications[7]. Under an external magnetic field, MNPs can generate stray fields. By integrating with various magnetic sensors such as magnetoresistance sensors[8,9], hall sensors[10,11], nuclear magnetic resonance (NMR) sensors[11], magnetic resonance imaging (MRI)[12], and magnetic particle spectroscopy (MPS)[13], MNPs can also serve as magnetic markers in diagnostic applications.

To date, MNPs with various sizes and surface coatings have been successfully commercialized and are available in many companies such as Ocean NanoTech (San Diego, USA), Nanoprobes (New York City, USA), US Research Nanomaterials (Houston, USA), Miltenyi Biotec (Bergisch Gladbach, Germany), micromod Partikeltechnologie GmbH (Rostock, Germany) and nanoComposix (San Diego, USA), etc. For those aforementioned applications, the quest for high magnetic moment, uniform size distribution, and colloidal stability MNPs has pushed the development of various nanoparticle manufacturers. In this paper, we first characterized the magnetic and physical properties of single-core, differently-sized iron oxide nanoparticle products from Ocean NanoTech using Vibrating-Sample Magnetometer (VSM), X-Ray Diffraction (XRD), Dynamic Light Scattering (DLS), Transmission Electron Microscopy (TEM), and Zeta Potential Analyzer (summarized in Table 1). In addition, we give application-oriented assessments on these MNP products using a home-built MPS system. Practical suggestions on the applications of these iron oxide nanoparticle with varying core sizes are given at the end of this paper to maximize the use of them.

## 2. MATERIALS AND METHODS

**2.1. Materials.** The SHA series MNPs are provided by the Ocean NanoTech. Six SHA series MNPs with average magnetic core sizes of 5 nm, 10 nm, 15 nm, 20 nm, 25 nm, and 30 nm are characterized in this paper (denoted as SHA-5, SHA-10, SHA-15, SHA-20, SHA-25, and SHA-30, respectively. Photographs of SHA series MNPs used in this work can be found in Supporting Information S1). The SHA series MNPs are a group of water-soluble iron oxide nanoparticles coated with amphiphilic polymer and functionalized amine reactive groups. They are very stable in most buffers in the pH range of 4 – 10 and can be readily conjugated to protein, peptide and other carboxylic acid containing molecules.

**2.2. Vibrating Sample Magnetometer (VSM) Measurement.** 25 uL of SHA series MNP suspension is pipetted onto filter paper and air-dried before the VSM measurements. Three independent magnetization curves of each sample are obtained at 20 °C, with the external magnetic field swept from -5000 to +5000 Oe (field step of 10 Oe and averaging time of 200 ms), -500 to +500 Oe (field step of 2 Oe and averaging time of 200 ms), and -200 to +200 Oe (field step of 1 Oe and averaging time of 200 ms), respectively.

**2.3. X-Ray Diffraction (XRD) Measurement.** 50 µL SHA series MNPs suspension is pipetted onto a Si/SiO$_2$ slide and air-dried before the XRD characterization. Cobalt radiation source (wavelength ~1.79 Å) is used for the XRD characterization since it has lower fluorescence especially for magnetite and maghemite[14]. For a convenient comparison, the characterized XRD patterns are converted to copper radiation. The crystal structure of SHA series MNPs are characterized via the x-ray diffraction (XRD, Bruker D8 Discover 2D).

**2.4. Dynamic Light Scattering (DLS) Measurement.** The hydrodynamic size distribution of the SHA series MNPs are characterized using DLS Particle Tracking Analyzer (Model: Microtac Nanoflex). 100 µL of the SHA series MNP suspension is diluted in 1.4 mL of DI water, reaching a total sample volume of 1.5 mL mixture and followed by ultra-sonication for 30 minutes before the DLS characterization.

**2.5. Transmission Electron Microscopy (TEM) Analysis.** The morphologies of these SHA series MNPs are characterized by a TEM system (FEI T12 120 kV). Each TEM sample is prepared by putting a droplet (~10 µL) of MNP suspension onto a TEM grid (copper mesh coated with amorphous carbon film). These samples are ready for TEM characterization when the solutions are fully evaporated at room temperature in air.

**2.6. Zeta Potential Measurement.** Zeta Potential Analyzer (Model: Stabino) is used to characterize the particle charge distribution or the zeta potential of the SHA series MNPs in DI water. 100 µL of SHA series MNP is diluted in 4.9 mL of DI water, reaching a total sample volume of 5 mL, followed by ultra-sonication for 30 minutes and then used for zeta potential characterization. This particle charge characterization helps analyze the surface binding capabilities of these SHA series MNPs.

**2.7. Magnetic Particle Spectroscopy (MPS) Measurement.** The dynamic magnetic responses of SHA series MNPs are characterized by a home built MPS system (see the schematic view and photographs of MPS system in S2 and S3 from Supporting Information). 200 µL of SHA series MNP sample is sealed in a plastic vial (maximum capacity of 300 µL). Two sets of copper coils are used to generate sinusoidal magnetic fields with tunable frequencies and magnitudes. One pair of differentially wound pick-up coils (600 windings clockwise and 600 windings counter-clockwise) collects the induced voltage signals due to the dynamic magnetic responses of MNPs under driving magnetic fields. A laptop with LabVIEW controls the frequency and magnitude of driving magnetic field through a data acquisition card (DAQ, NI USB-6289). The analog voltage signals are sent back from pick-up coils to DAQ, sampled at 500 kHz, and converted to frequency domain after discrete Fourier transform (DFT). For each MPS measurement, the MPS system runs for 10 seconds to collect the baseline signal (noise) followed by inserting the vial containing MNP sample for another 10 seconds' signal (total) collection. The induced voltage due to dynamic magnetic responses of MNPs is recovered from the total signal by the phasor theory (see S4 from Supporting Information). The higher harmonics specific to dynamic magnetic responses of MNPs are extracted for analysis (see S7 from Supporting Information).

3. **RESULTS AND DISCUSSION**

**3.1. Magnetic Properties of SHA Series MNPs.** The hysteresis curves of SHA series MNPs are recorded by VSM, under field ranges of -5000 Oe – 5000 Oe, -500 Oe – 500 Oe, and -200 Oe – 200 Oe. The magnetic moment per microliter is averaged over 25 µL MNP sample and plotted in Figure 1(a) – (c). Under field strength of 500

Oe, the saturation magnetizations from highest to lowest are: SHA-15 > SHA-20 > SHA-30 > SHA-25 > SHA-10 > SHA-5. With SHA-5, SHA-10, SHA-15, and SHA-20 being superparamagnetic. We also observed hysteresis loops from SHA-25 and SHA-30. The SHA-5 MNPs show very large coercivity, which may due to the surface spin-canting effect [15]. Due to the varying particle concentrations in SHA series MNP products (listed in Table 1), the magnetic moment per microliter is not comprehensive to represent the magnetic property of each MNP. In addition, magnetic moment per particle is also summarized in Figure 1(d) – (f), with SHA-30 showing the highest magnetic moment per particle followed by SHA-25, SHA-20, SHA-15, SHA-10, and SHA-5 showing the lowest magnetic moment per particle.

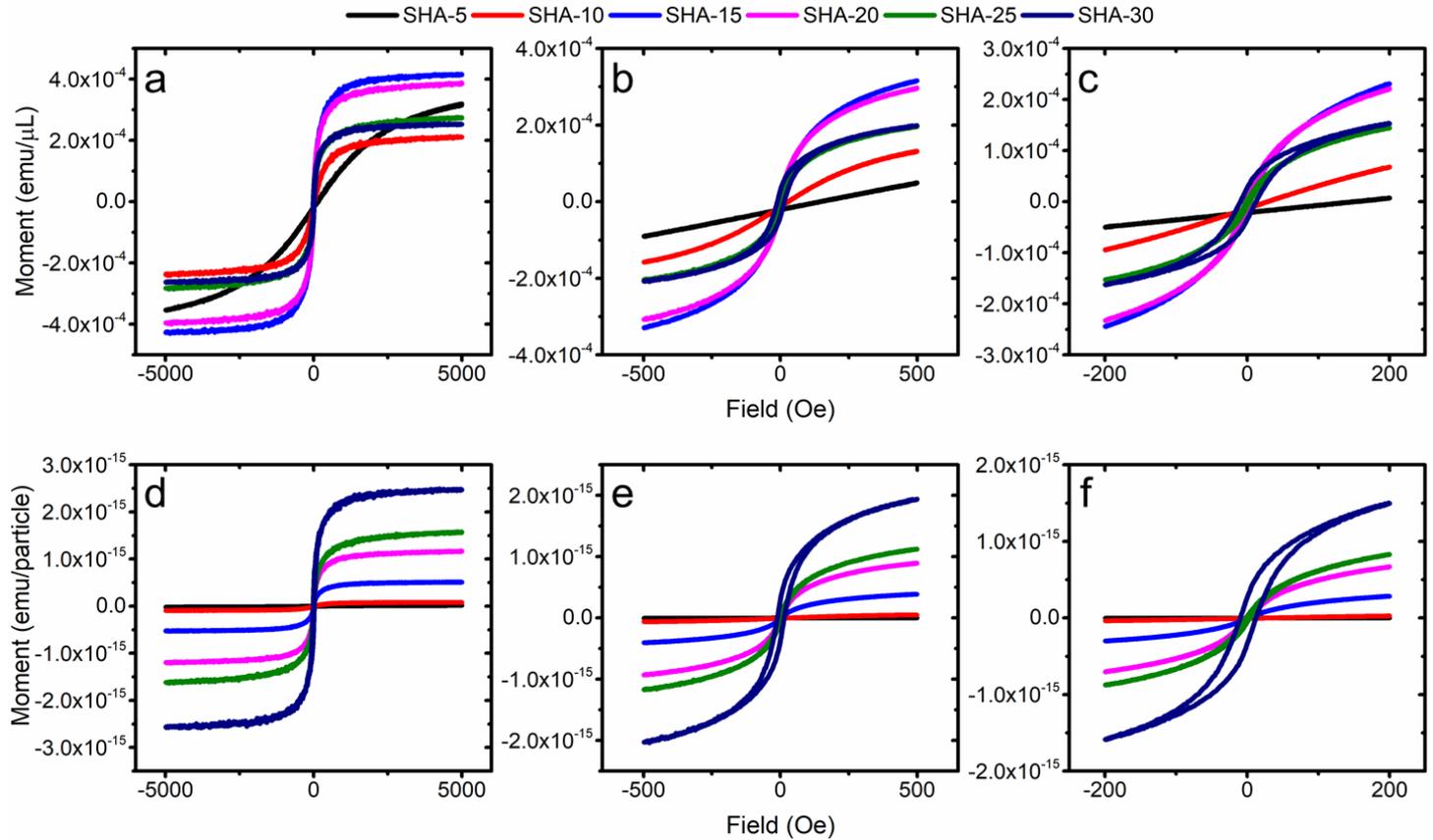

Figure 1. Magnetization curves of SHA series MNPs obtained by VSM at 20 °C. External field sweeps from (a) & (d) -5000 to +5000 Oe, (b) & (e) -500 to +500 Oe, and (c) & (f) -200 to +200 Oe. Magnetization units are represented by emu/μL and emu/particle for (a) – (c) and (d) – (f), respectively.

The crystal structure of SHA series MNPs are characterized via the x-ray diffraction (XRD, Bruker D8 Discover 2D) as shown in Figure 2. It is observed that $Fe_3O_4$ and $\gamma$-$Fe_2O_3$ are the main phases in SHA series MNPs. There are also several diffraction peaks from the solution denoted by the blue dashed lines in Figure 2. The sharp diffraction peaks (labeled by black diamonds) come from the chemicals in the MNP buffer (NaCl, KCl, $Na_2HPO_4$, $KH_2PO_4$, etc), and the peaks labeled by black rounds come from the Si/$SiO_2$ substrate. The full width at half maximum (FWHM) of the diffraction peaks are wider for the MNPs compared to their bulk counterparts. This is due to the peak broadening effects for the nanoscale materials based on Scherrer equation.

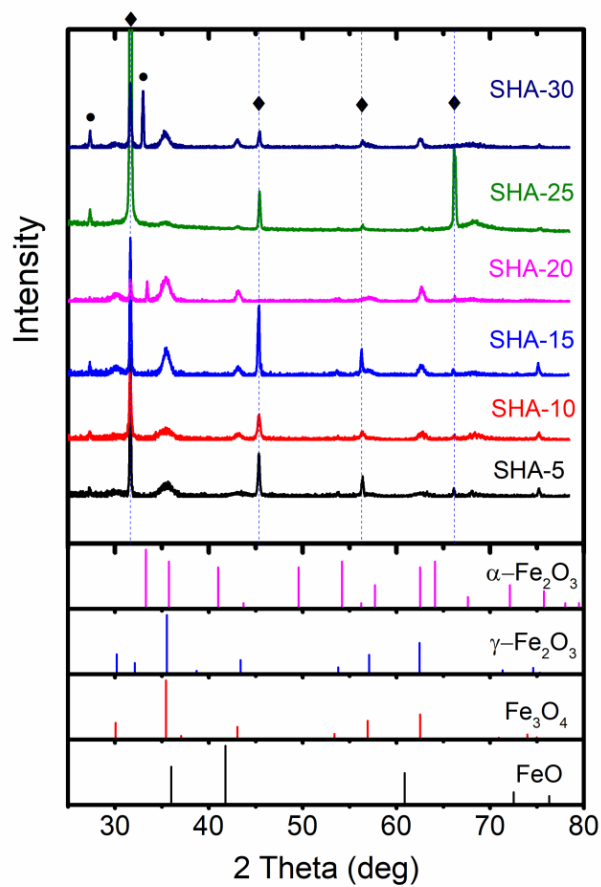

Figure 2. XRD patterns of SHA series MNPs. The powder diffraction files (PDFs) of FeO, $Fe_3O_4$, $\alpha$-$Fe_2O_3$, and $\gamma$-$Fe_2O_3$ are added for references.

Table 1. Physical and magnetic properties of SHA series MNPs.

| Sample | Average Size (nm) ± SD | Zeta Potential (mV) ± SD | Saturation Magnetization (emu/g Fe)[a] | Saturation Magnetization (emu/g)[b] | Magnetic Moment (emu/particle) | Material |
|---|---|---|---|---|---|---|
| **SHA-5** | 10.46 ± 3.88 | -0.03 ± 0.005 | 63.84 | ~44.69 | $1.54 \times 10^{-17}$ | $\gamma$-$Fe_2O_3$, $Fe_3O_4$ |
| **SHA-10** | 18.07 ± 4.72 | 5.03 ± 0.07 | 42.64 | ~29.85 | $8.23 \times 10^{-17}$ | $\gamma$-$Fe_2O_3$, $Fe_3O_4$ |
| **SHA-15** | 20.69 ± 6.31 | 7.66 ± 0.05 | 83.44 | ~58.41 | $5.13 \times 10^{-16}$ | $\gamma$-$Fe_2O_3$, $Fe_3O_4$ |
| **SHA-20** | 27.56 ± 11.29 | -0.41 ± 0.33 | 78.08 | ~54.66 | $1.18 \times 10^{-15}$ | $\gamma$-$Fe_2O_3$, $Fe_3O_4$ |
| **SHA-25** | 28.28 ± 10.38 | 1.15 ± 0.49 | 55.28 | ~38.70 | $1.58 \times 10^{-15}$ | $\gamma$-$Fe_2O_3$, $Fe_3O_4$ |
| **SHA-30** | 32.60 ± 12.17 | -0.69 ± 0.04 | 51.12 | ~35.78 | $2.50 \times 10^{-15}$ | $\gamma$-$Fe_2O_3$, $Fe_3O_4$ |

[a] Magnetic moment per gram of Fe (emu/g Fe) vs. field can be found in S5 from Supporting Information.
[b] Magnetic moment per gram of iron oxide nanoparticle is calculated by assuming iron holds 70% of nanoparticle weight.

**3.2. Hydrodynamic Size and Morphological Characterizations on SHA Series MNPs.** Figure 3(a) – (f) shows the hydrodynamic size distributions of samples SHA-5, SHA-10, SHA-15, SHA-20, SHA-25 and SHA-30 with mean values of 10.46 nm, 18.07 nm, 20.69 nm, 27.56 nm, 28.28 nm and 32.6 nm, respectively. The SHA series MNPs have two organic coating layers: one monolayer of oleic acid and another monolayer of amphiphilic polymer. The total thickness of the organic layer coating is about 4 nm. This causes the hydrodynamic size of the nanoparticles to be about 8 - 10 nm larger than their inorganic core size measured by TEM (see Figure 4). In this respect, the mean values for the DLS in the SHA series show quite a satisfactory trend and good agreement with the TEM results; the results are arranged in ascending order of their sizes in the Figure 3.

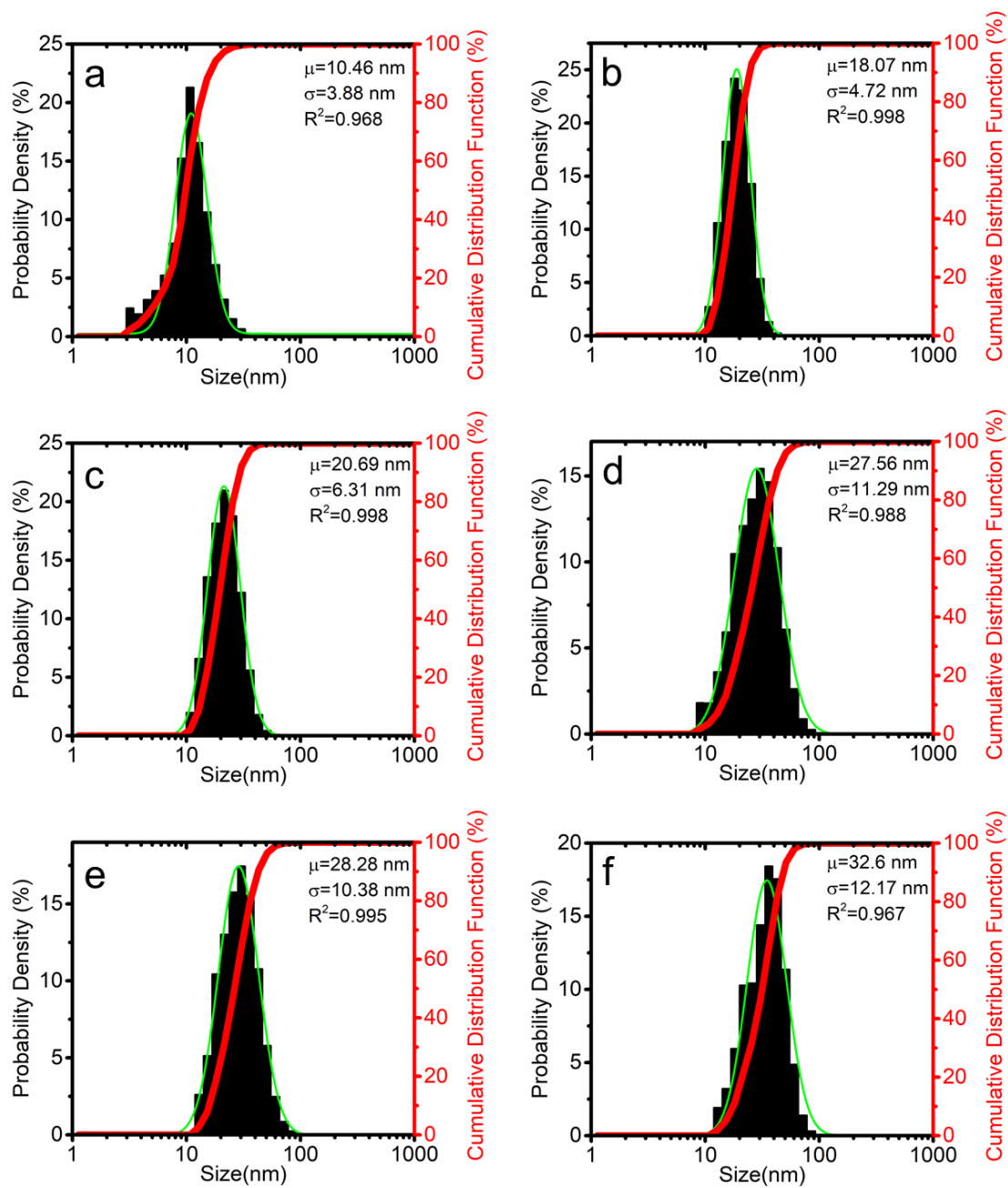

Figure 3. Statistical distributions of the hydrodynamic size of the samples (a) SHA-5, (b) SHA-10, (c) SHA-15, (d) SHA-20, (e) SHA-25 and (e) SHA-30 as characterized by DLS. In each figure, the solid green lines are the fitted log-normal distribution curves and the solid red lines are the cumulative distribution curves. µ values represent the statistical mean of the hydrodynamic sizes of the samples. The standard deviation and R-square values are represented by σ and $R^2$, respectively for each case.

The magnetic core morphologies of SHA series MNPs are shown in Figure 4. Some MNPs are agglomerated during the evaporation process of the MNP suspensions. For the MNPs with smaller size such as the samples SHA-5 and SHA-10, the magnetic core shapes are irregular. However, larger MNPs show spherical magnetic cores. The contrast of different MNPs from one TEM image is due to the different crystal orientations. When the crystal zone axis is close to the incident electron beam, the MNPs show darker color.

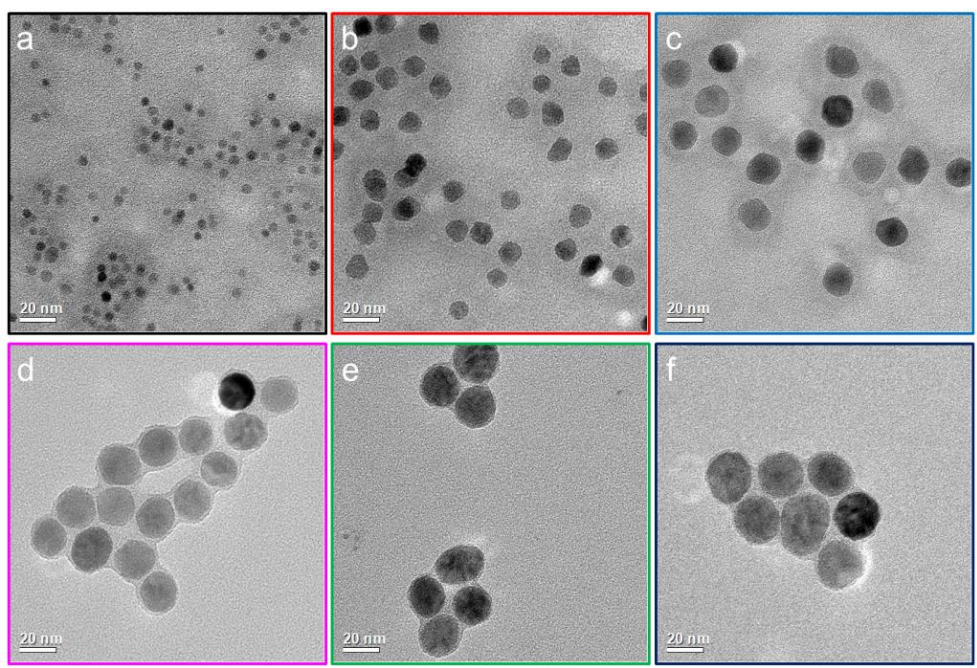

Figure 4. The TEM images of SHA series MNPs. (a) - (f) corresponds to SHA-5, SHA-10, SHA-15, SHA-20, SHA-25, and SHA-30, respectively. Scale bars represent 20 nm. TEM images of SHA series MNPs under different magnifications are given in S6 from Supporting Information.

**3.3. Zeta Potential of SHA Series MNPs.** The SHA series MNPs have a neutral to slightly alkaline pH between 7.2 and 7.6. The measured zeta potential values for SHA-5, SHA-10, SHA-15, SHA-20, SHA-25 and SHA-30 are -0.03 mV, +5.03 mV, +7.66 mV, -0.41 mV, +1.15 mV and -0.69 mV, respectively.

**3.4. Dynamic Magnetic Responses of SHA Series MNPs under Low Frequency Driving Field.** The dynamic magnetic responses of SHA series MNPs under mono-frequency driving field are investigated. The driving field frequency is varied from 50 Hz to 2850 Hz and the field amplitude is set at 170 Oe (Gauss).[16,16–23] Each plastic vial containing 200 µL SHA series MNPs in 10 nM PBS and 0.03% $NaN_3$ is placed under the alternating magnetic field for MPS measurements. For MNPs suspended in liquid solution under external magnetic field, they undergo two distinct relaxation mechanisms by which the magnetic moments rotate in response to the field: the Néel relaxation is the rotation of magnetic moment inside a stationary MNP and, on the other hand, the Brownian relaxation is the physical rotation of the entire MNP along with its magnetic moment. In principle, both relaxation mechanisms play important roles in determining the dynamic magnetic responses of MNPs in suspension when subjected to alternating magnetic field. Depending on the magnetic properties (such as effective anisotropy constant, saturation magnetization) [24,25], the physical properties (magnetic core size, the hydrodynamic size including the polymer coatings and anchored biological compounds such as protein, peptide, cells, etc.) of MNPs [15,17,26–31], the nanoparticle volume fraction of suspension (i.e., dipolar interactions) [15,32–34], the physical properties of the suspension (temperature, viscosity) [18,19,35–42], MNPs could undergo either Néel or Brownian process-dominated relaxation. It has been reported that for a system of non-interacting iron oxide nanoparticles with negligible polymer coatings, the magnetic dynamics will be dominated by Brownian process when the core size is above 15 nm and Néel process dominates when the core size is below 15 nm [35,43–45] (see S8 from Supporting Information).

Under low frequency driving field (f < 500 Hz), magnetic moments of SHA series MNPs with diameters from 5 nm to 30 nm are able to follow the time-varying magnetic field. As shown in Figure S9 from Supporting Information, all of the six SHA series MNPs show similar phase angles to the driving field (f < 500 Hz), as the field frequency increases, the differences in the phase angles between six samples increase. Larger MNPs with larger effective relaxation time show larger phase lag to the driving field.

As is summarized in Figure 5, under low frequency driving field, the dynamic magnetic responses of six SHA series MNPs from strongest to weakest are: SHA-30 > SHA-20 > SHA-15 > SHA-25 > SHA-10 > SHA-5. Figure 5(a), (b) & (c) summarize the amplitudes measured at the $3^{rd}$, the $5^{th}$, and the $7^{th}$ harmonics, respectively. Figure 5(d), (e) & (f) highlight the corresponding harmonic amplitudes under driving field frequencies of 350 Hz, 650 Hz, 1250 Hz, and 1850 Hz.

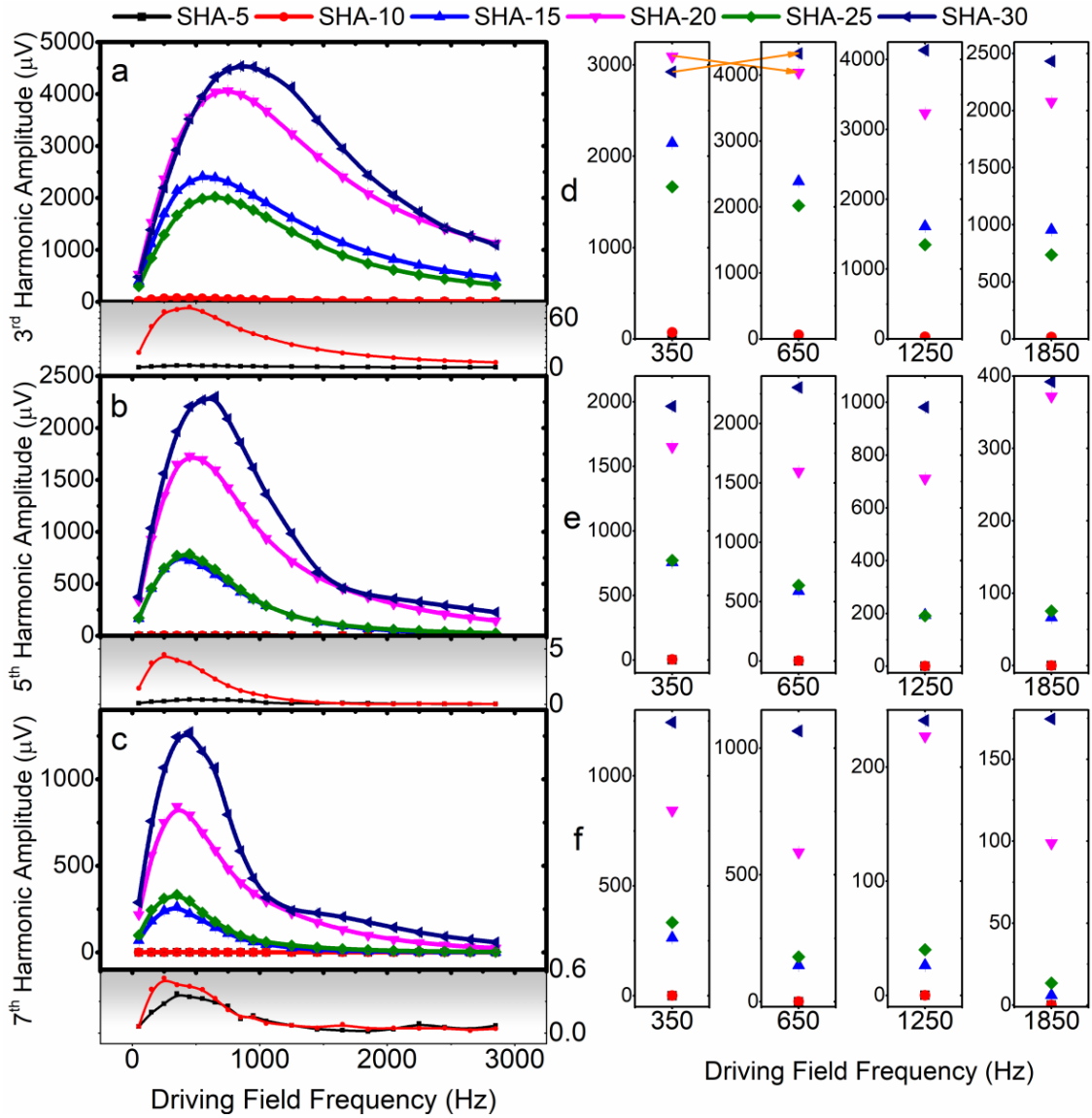

Figure 5. Harmonics generated by SHA series MNPs under low frequency driving field. (a) – (c) summarize the 3$^{rd}$, the 5$^{th}$, and the 7$^{th}$ harmonic amplitudes of SHA series MNPs under different driving field frequencies. (d) – (f) highlight the harmonic amplitudes at driving field frequency of 350 Hz, 650 Hz, 1250 Hz, and 1850 Hz.

Figure 6 summarizes the real-time voltage signal collected from pick-up coils at driving field frequencies of 350, 950, and 1850 Hz. The extracted harmonics are plotted along with the total signal in real time. MNPs with stronger dynamic magnetic responses to the driving field generate larger harmonic signals, thus are able to cause the distortions in the total signal (the highlighted dark areas in Figure 6). It is observed that SHA-30 and SHA-20 show the strongest dynamic magnetic responses to the low frequency driving field, followed by SHA-15 and SHA-25. SHA-5 and SHA-10 shows negligible dynamic magnetic responses compared to the former SHA series MNPs. Which is mainly due to the low magnetic moments and linear magnetization curves as show in Figure 1.

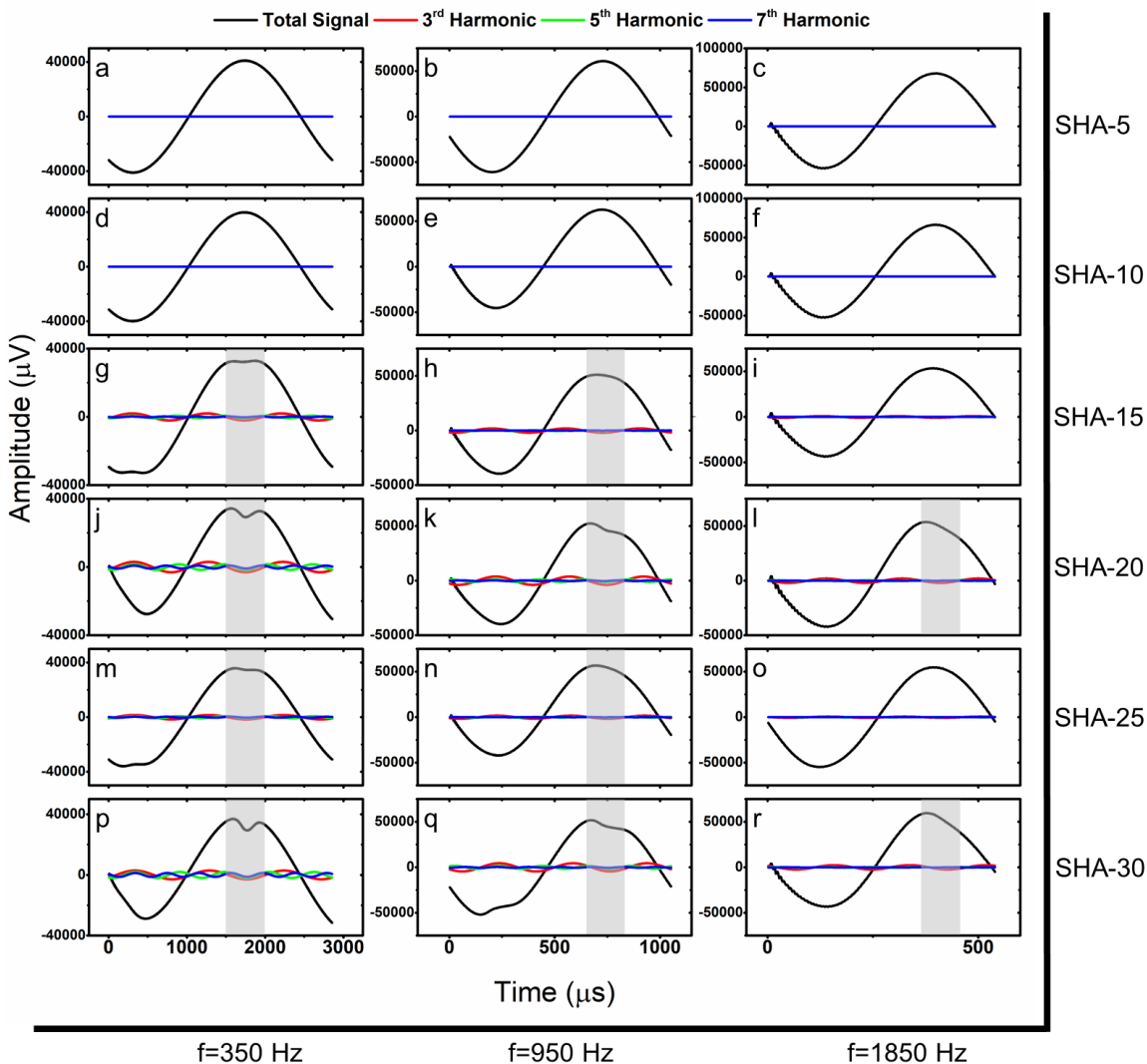

Figure 6. Recorded real-time dynamic magnetic responses of SHA series MNPs under low frequency driving field. The higher harmonics are extracted and plotted in parallel with the total signal collected from the pick-up coils.

**3.5. Dynamic Magnetic Responses of SHA Series MNPs under High Frequency Driving Field.** In this section, we report the dynamic magnetic responses of SHA series MNPs under high frequency driving fields. A dual-frequency method is used herein, one excitation field is set at 10 Hz and magnitude of 170 Oe, the other high frequency driving field is set at varying frequencies (from 1 kHz to 20 kHz) and magnitude of 17 Oe.[13,26,35,43,46–48]

Under high frequency driving field, larger MNPs (i.e., SHA-30) are unable to rotate their magnetic moments to the fast-switching magnetic field, thus, their dynamic magnetic responses are weakened. As shown in Figure S10 from Supporting Information, there is a constant difference of 50° harmonic phase differences between SHA-10 and SHA-30 MNPs. As a result, under high frequency driving field, the dynamic magnetic responses of six

SHA series MNPs from strongest to weakest are: SHA-15 > SHA-20 > SHA-30 > SHA-25 > SHA-10 > SHA-5 (as shown in Figure 7).

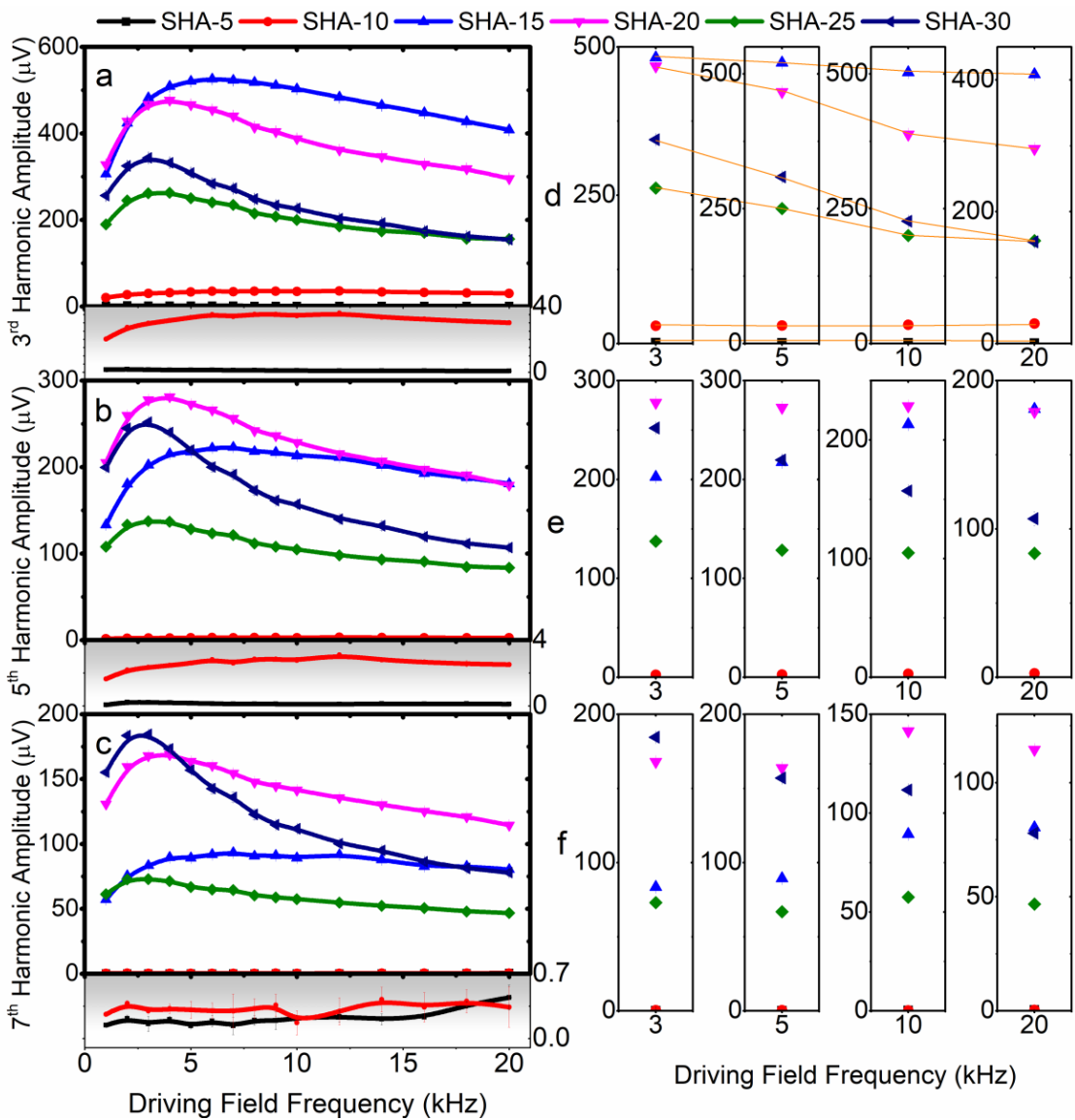

Figure 7. Harmonics generated by SHA series MNPs under high frequency driving field. (a) – (c) summarize the 3$^{rd}$, the 5$^{th}$, and the 7$^{th}$ harmonic amplitudes of SHA series MNPs under different driving field frequencies. (d) – (f) highlight the harmonic amplitudes at driving field frequency of 3 kHz, 5 kHz, 10 kHz, and 20 kHz.

Although recent in origin, MNPs of different core sizes have found their applications in various fields of science. This section of the paper is dedicated in identifying the utility of the different sized and surface functionalized MNPs in realistic applications. The SHA series particles are amine functionalized MNPs. As the amine groups are less selective and less specific for antibodies and proteins, they capture a varied range of bacterial pathogens and allow purification of water, food and urine samples [49]. The VSM characterization of the SHA series in Figure 1(a) - (f) show that SHA-5, SHA-10, SHA-15 and SHA-20 are superparamagnetic. Although SHA-25 & SHA-30 show higher magnetic moments, they show hysteresis loops. For magnetic biosensing, higher moment particles are preferred in order to generate higher magnetic signal per particle. In the meantime, the MNPs are required to be superparamagnetic to prevent aggregations. For the SHA series, SHA-25 exhibits the second highest magnetic moment/particle with a remanent magnetization of 1.28%$M_s$, where $M_s$ is the saturation magnetization. Although SHA-30 has a higher magnetic moment/particle compared to SHA-25, a much larger remanence magnetization of 10.93%$M_s$ is observed from SHA-30. Taking both magnetic moments and superparamagnetism into consideration, SHA-25 is the optimum candidate from SHA series for biosensing applications. On a different note, for cell separation & sorting and drug/gene delivery, as the property of superparamagnetism is not essential and higher magnetic moment ensures larger magnetic torque (force), the highest magnetic moment MNP, SHA-30 is probably a better candidate. For magnetic hyperthermia therapy, the area of magnetic hysteresis loop, $A$, corresponds to the dissipated energy or specific absorption rate (SAR), which is evaluated by the equation $SAR = A \cdot f$. Since the maximum SAR achievable is directly proportional to the $M_s$ of MNPs [50–54]. Hence the SHA-30 MNPs are best for hyperthermia treatments. Furthermore, magnetic resonance imaging (MRI) techniques require the MNPs to be injected into the body fluids which then accumulate in the target tissues. Hence, for MRI applications it is extremely essential for the MNPs to be small as larger MNPs have greater tendency to block the arteries. In that case, SHA-5 & SHA-10 MNPs will be quite useful [55]. The dynamic magnetic responses of SHA series MNPs are compared in this paper using a home-built MPS system. The harmonics are induced under different driving magnetic fields, which is a result of the joint effects of relaxation mechanisms and the magnetic moment of each MNP. For MPI and MPS-based bioassays, larger dynamic magnetic responses (higher harmonic amplitudes) ensure higher signal-to-noise ratio and sensitivity. Thus, SHA-30 MNPs are suggested for MPI and MPS-based bioassays where the driving field frequencies are below 2 kHz, while SHA-15 MNPs are suggested for these applications where the driving field frequencies are above 2 kHz.

## 4. CONCLUSIONS

In this paper, we characterized the magnetic and physical properties of SHA series MNPs from Ocean NanoTech, using standard characterization tools. The VSM results show that SHA-5, SHA-10, SHA-15, and SHA-20 MNPs are superparamagnetic and, on the other hand, SHA-25 and SHA-30 are not superparamagnetic. With SHA-30 showing the highest magnetic moment per particle, followed by SHA-25, SHA-20, SHA-15, SHA-10, and SHA-5. Thus, SHA series iron oxide nanoparticles with larger core sizes are preferred for magnetic biosensing and drug delivery where high moment MNPs are desired for higher magnetic signals and higher magnetic torques. However, SHA-25 and SHA-30 show remnant magnetizations upon the removal of magnetic field (non-superparamagnetic), thus they are not applicable for applications where superparamagnetism is required. The XRD results show that all SHA series MNPs are composed of γ-$Fe_2O_3$, $Fe_3O_4$. The dynamic magnetic responses of these iron oxide nanoparticles are investigated by a home-built MPS system, where both the responses under low and high driving field frequencies are summarized. It is observed that under low driving field frequencies, the dynamic magnetic responses of SHA series MNPs from strongest to weakest are: SHA-30 > SHA-15 > SHA-25 > SHA-10 > SHA-5. However, under high driving field frequencies, due to the larger phase lags of larger MNPs, the dynamic magnetic responses from strongest to weakest are modified: SHA-15 > SHA-20 > SHA-30> SHA-25 > SHA-10 > SHA-5. These results give hints on designing MPI and MPS-based bioassays to maximize the use of different MNPs of different core sizes. At the end of this paper, based on the requirements and goals of MNP-based applications, we suggested different SHA MNPs for each application.

## ASSOCIATED CONTENT

**Supporting Information**

Photographs of SHA series iron oxide nanoparticles; Magnetic Particle Spectroscopy (MPS) system; Magnetic Particle Spectroscopy (MPS) system setups; Phasor theory; Magnetic moment per gram of Fe; TEM images of SHA series MNPs captured at various magnifications; Magnetic dynamic responses and higher harmonic models; Brownian and Néel relaxations; Phase angles of higher harmonics monitored under low frequency driving field for SHA series samples; Phase angles of higher harmonics monitored under high frequency driving field for SHA series samples.

## AUTHOR INFORMATION

**Corresponding Authors**

*E-mail: wuxx0803@umn.edu (K.W.)

*E-mail: jpwang@umn.edu (J.-P.W.)

**ORCID**

Kai Wu: 0000-0002-9444-6112

Jinming Liu: 0000-0002-4313-5816

Renata Saha: 0000-0002-0389-0083

Chaoyi Peng: 0000-0003-1608-3886

Diqing Su: 0000-0002-5790-8744

Jian-Ping Wang: 0000-0003-2815-6624

**Author Contributions**

[⊥]K.W., J.L., and R.S. contributed equally to this work.

**Notes**

The authors declare no conflict of interest.

## ACKNOWLEDGMENTS

This study was financially supported by the Institute of Engineering in Medicine of the University of Minnesota through FY18 IEM Seed Grant Funding Program. Portions of this work were conducted in the Minnesota Nano Center, which is supported by the National Science Foundation through the National Nano Coordinated Infrastructure Network (NNCI) under Award Number ECCS-1542202. Portions of this work were carried out in the Characterization Facility, University of Minnesota, a member of the NSF-funded Materials Research Facilities Network (www.mrfn.org) via the MRSEC program.

**TOC Graphic**

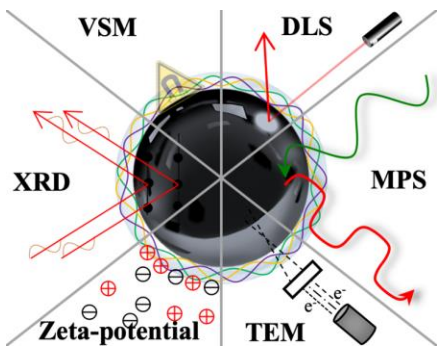

# An Investigation of Commercial Iron Oxide Nanoparticles: Advanced Structural and Magnetic Properties Characterization

Kai Wu[†,⊥,*], Jinming Liu[†,⊥], Renata Saha[†,⊥], Chaoyi Peng[†], Diqing Su[‡], Andrew Yongqiang Wang[§,*], and Jian-Ping Wang[†,*]

[†]Department of Electrical and Computer Engineering, University of Minnesota, Minneapolis, Minnesota 55455, USA

[‡]Department of Chemical Engineering and Material Science, University of Minnesota, Minneapolis, Minnesota 55455, USA

[§]Ocean NanoTech, LLC, San Diego, California 92126, USA

*Correspondence and requests for materials should be addressed to K.W. (email: wuxx0803@umn.edu), A.W. (email: awang@oceannanotech.com), and J.-P.W. (email: jpwang@umn.edu).

[⊥]These authors contributed equally to this work.

**S1. Photographs of SHA series iron oxide nanoparticles.**

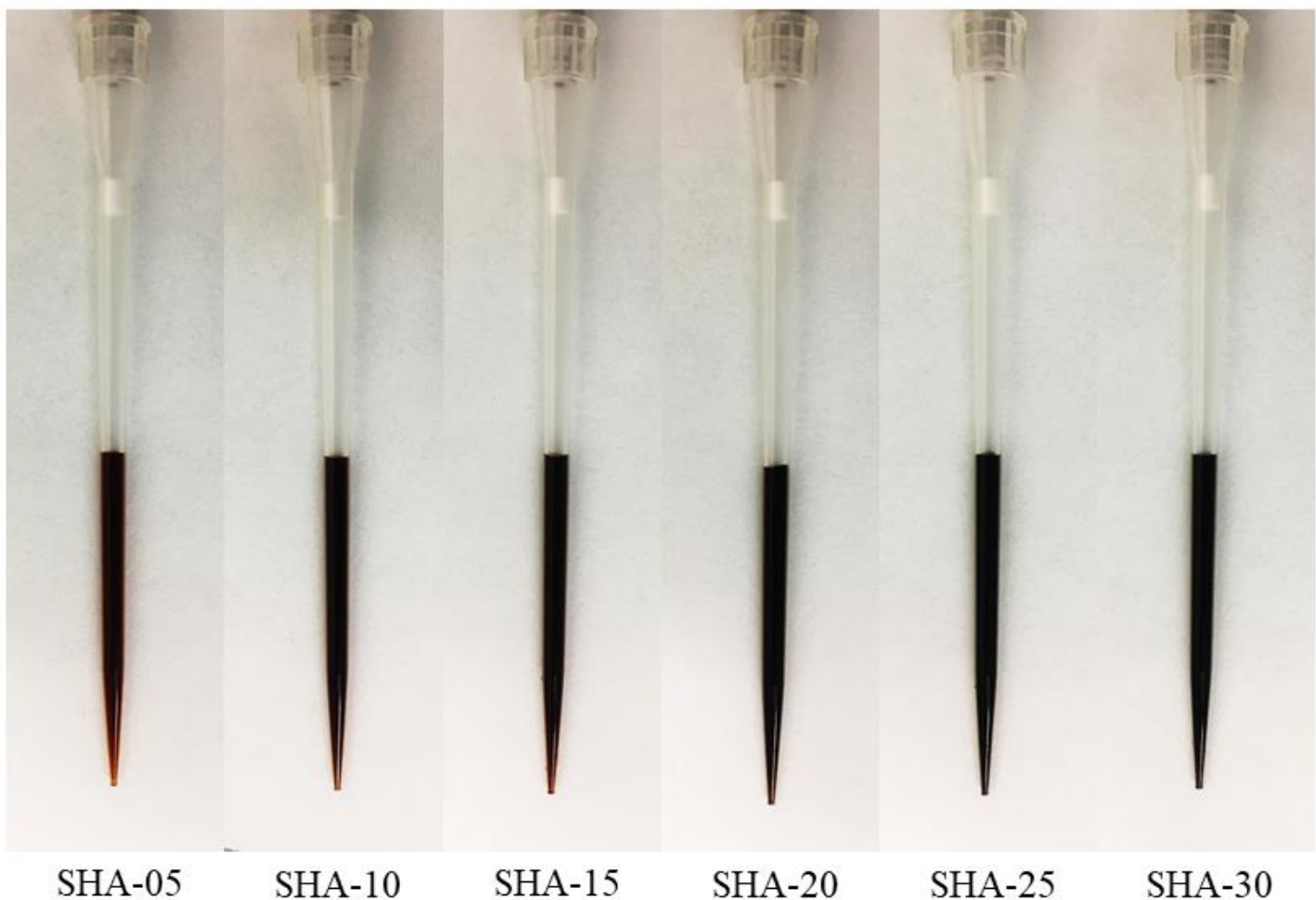

Figure S1. Photographs of SHA series iron oxide nanoparticles. Each pipette tip contains 100 µL liquid sample of SHA-X (X=5, 10, 15, 20, 25, and 30). The color gradually changes from dark brown to dark black (left to right) with the increase of nanoparticle size.

## S2. Magnetic Particle Spectroscopy (MPS) system.

Since the pioneering work of Gleich and Weizenecker in 2005, magnetic particle imaging (MPI) has emerged as a new 3D tomographic technique for clinical diagnosis, vascular imaging, and therapy[1–10]. Unlike the nanoparticle-enhanced magnetic resonance imaging (MRI) where MNPs are used as supportive contrast agents, the MPI exploits the magnetic response signals directly from MNPs and thus the only visualized elements[11–16]. Magnetic particle spectroscopy (MPS)-based bioassay, a derivative technique of MPI, has gained a lot of attention in the area of magnetic immunoassays in recent years[10,17–27]. The MPS-based bioassay can be interpreted as a 0D MPI scanner where a sinusoidal magnetic field is applied to MNPs, which periodically drives MNPs into magnetically saturated region. The nonlinear magnetic responses of MNPs are monitored by a pair of specially designed pick-up coils and the frequency domain signals are extracted. Those higher harmonics that are specific to MNPs are interpreted as indicators of the binding status of MNPs to target analytes[19,20,22,28–30].

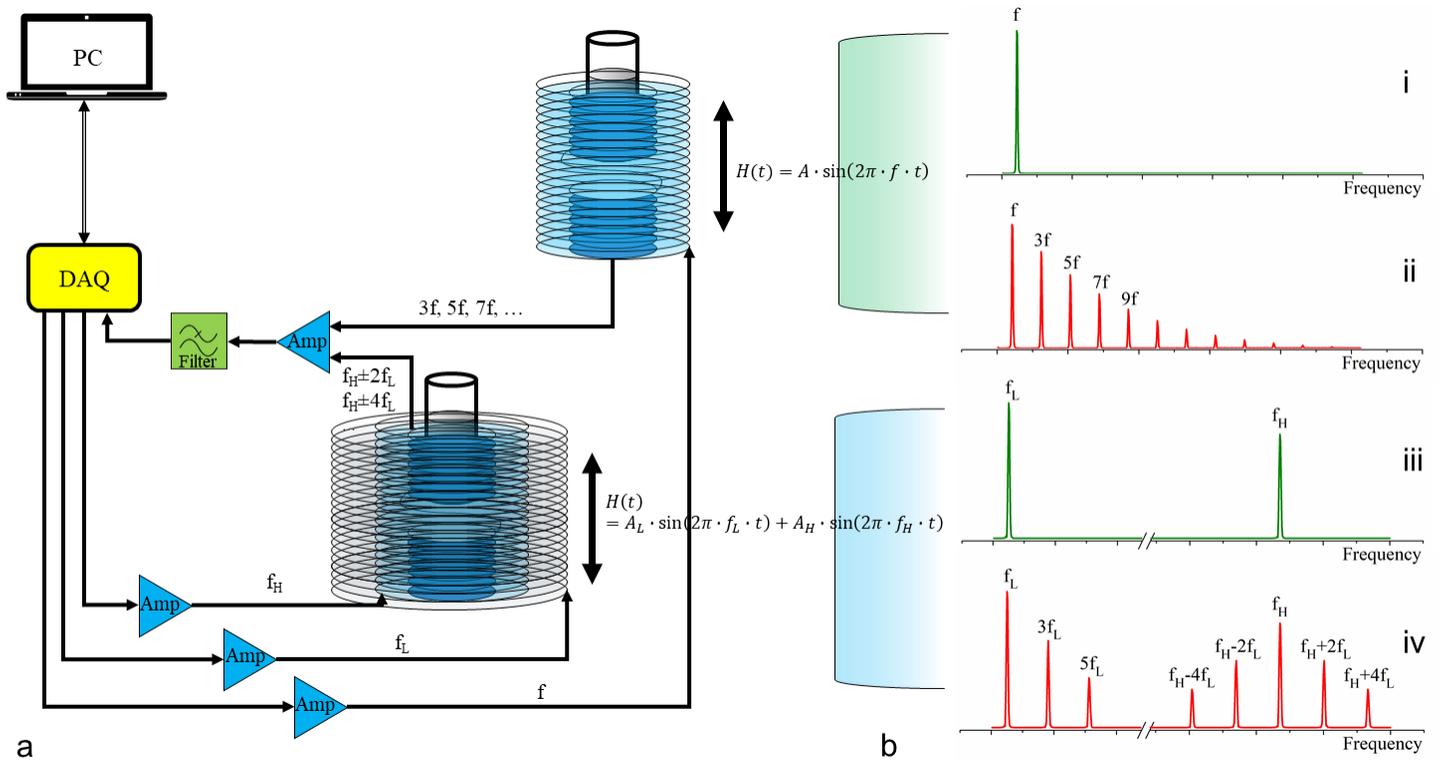

Figure S2. (a) Schematic view of mono- and dual-driving field MPS system. (b) Frequency domain representations of driving fields and magnetic responses of MNPs: i) mono-frequency driving field; ii) dynamic magnetic responses of MNPs to the mono-frequency field are picked up by the MPS system, higher harmonics containing 3f (denoted as the 3$^{rd}$ harmonic in this work), 5f (denoted as the 5$^{th}$ harmonic in this work), 7f (denoted as the 7$^{th}$ harmonic in this work), … are observed due to the nonlinear magnetic responses of MNPs; iii) dual-frequency driving field; iv) dynamic magnetic responses of MNPs to the dual-frequency fields are picked up by the MPS system, higher harmonics containing $3f_L$, $5f_L$, $7f_L$, …, $f_H \pm 2f_L$ (denoted as the 3$^{rd}$ harmonic in this work), $f_H \pm 4f_L$ (denoted as the 5$^{th}$ harmonic in this work), $f_H \pm 6f_L$ (denoted as the 7$^{th}$ harmonic in this work), …, $3f_H$, $5f_H$, $7f_H$, … are observed due to the nonlinear magnetic responses of MNPs.

**S3.** Magnetic Particle Spectroscopy (MPS) system setups.

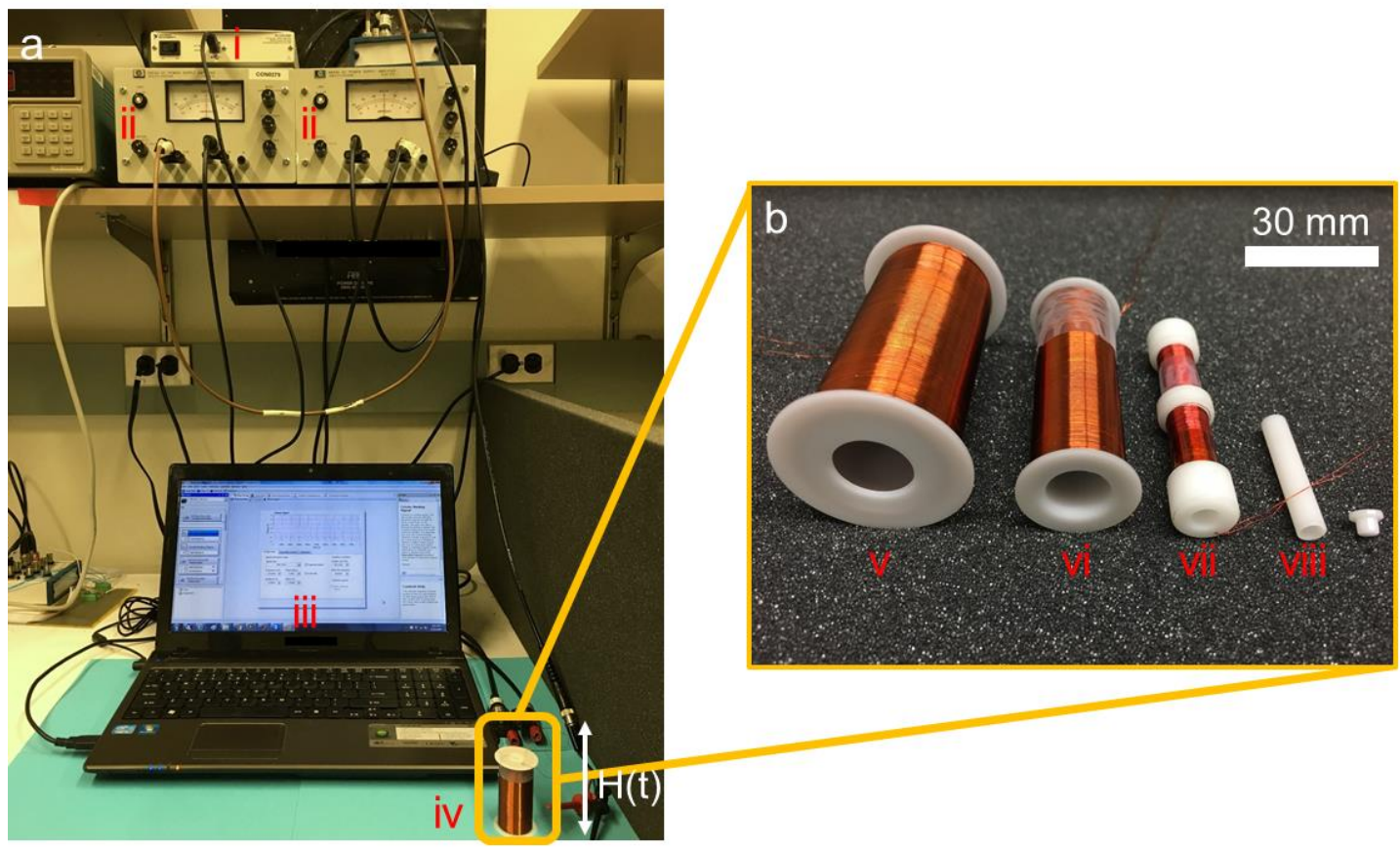

Figure S3. (a) Photograph of home-built MPS system setup. (b) Photograph of copper coils that generate driving magnetic fields and pick-up coils for signal collection. (i) DAQ; (ii) Amplifier; (iii) PC with LabVIEW; (iv) One set of assembled coils; (v)&(vi) are coils for the generation of driving magnetic fields; (vii) One pair of differentially wound pick-up coils; (viii) Plastic vial with maximum volume capacity of 300 µL.

## S4. Phasor theory.

The voltage and phase generated from MNPs at specific frequencies are represented by a phasor: $A \cdot e^{j(\omega t + \varphi)}$ (or expressed as $A \angle \varphi$), where $\omega$ is the angular frequency of driving field, $\varphi$ is the phase lag, and $j = \sqrt{-1}$.

For each MPS measurement, the MPS system is run for 10 seconds to collect the baseline signal (noise) followed by inserting the vial containing MNP sample for another 10 seconds' signal (total) collection. The background noise can be expressed as $A_{Noise} e^{j\varphi_{Noise}}$. The total signal is expressed as $A_{TOT} e^{j\varphi_{TOT}}$. This signal is the sum of two phasors: the background noise and the signal generated by MNPs (namely, $A_{MNP} e^{j\varphi_{MNP}}$).

So,

$$A_{Noise} e^{j\varphi_{Noise}} + A_{MNP} e^{j\varphi_{MNP}} = A_{TOT} e^{j\varphi_{TOT}},$$

which reduces to an equation set:

$$\begin{cases} A_{Noise} \times cos\varphi_{Noise} + A_{MNP} \times cos\varphi_{MNP} = A_{TOT} \times cos\varphi_{TOT} \\ A_{Noise} \times sin\varphi_{Noise} + A_{MNP} \times sin\varphi_{MNP} = A_{TOT} \times sin\varphi_{TOT} \end{cases}$$

By solving the equation set above, we can get the harmonic amplitude $A_{MNP}$ and phase $\varphi_{MNP}$ of each type of MNPs at different driving field frequencies.

## S5. Magnetic moment per gram of Fe.

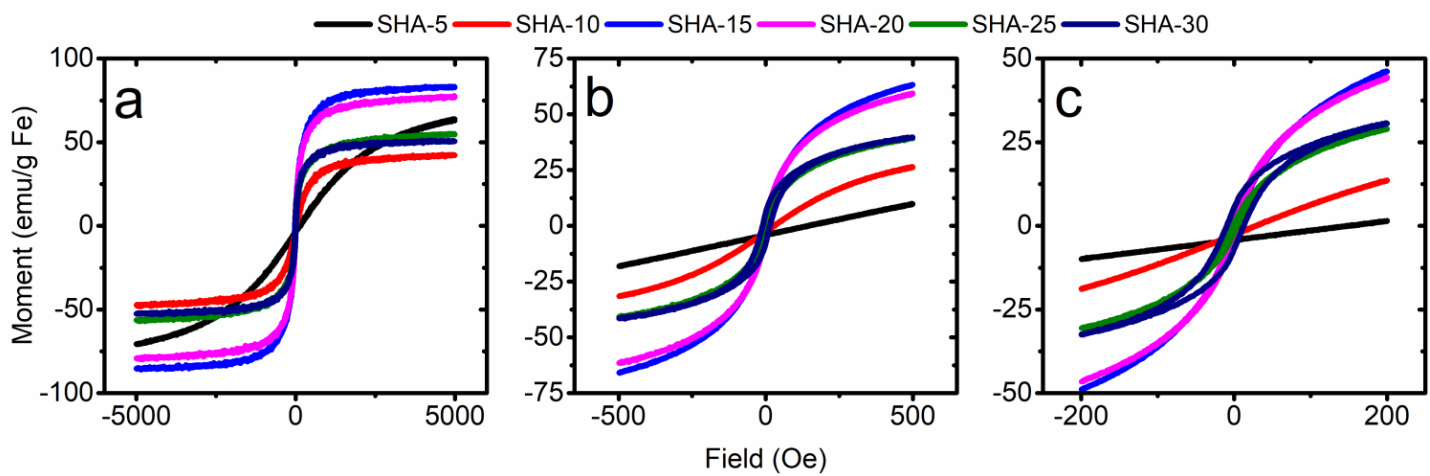

Figure S5. Magnetization curves of SHA series MNPs obtained by VSM.

**S6.** TEM images of SHA series MNPs captured at various magnifications.

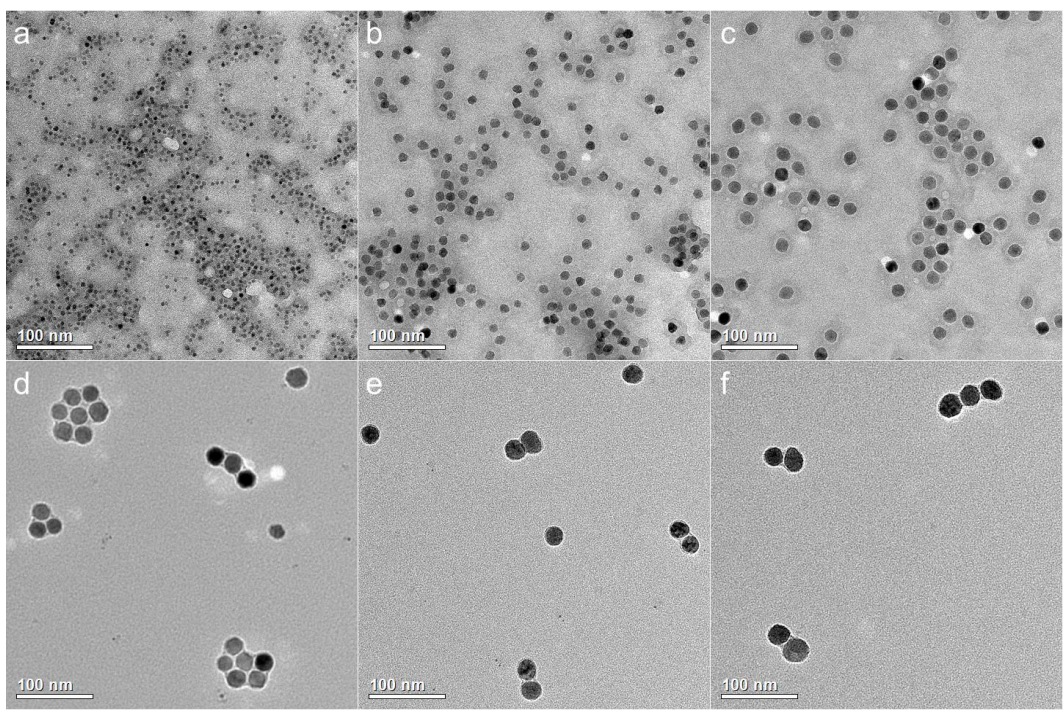

Figure S6-1. TEM images of SHA series MNPs. Scale bar represents 100 nm. (a) – (f) are SHA-5, SHA-10, SHA-15, SHA-20, SHA-25, and SHA-30, respectively.

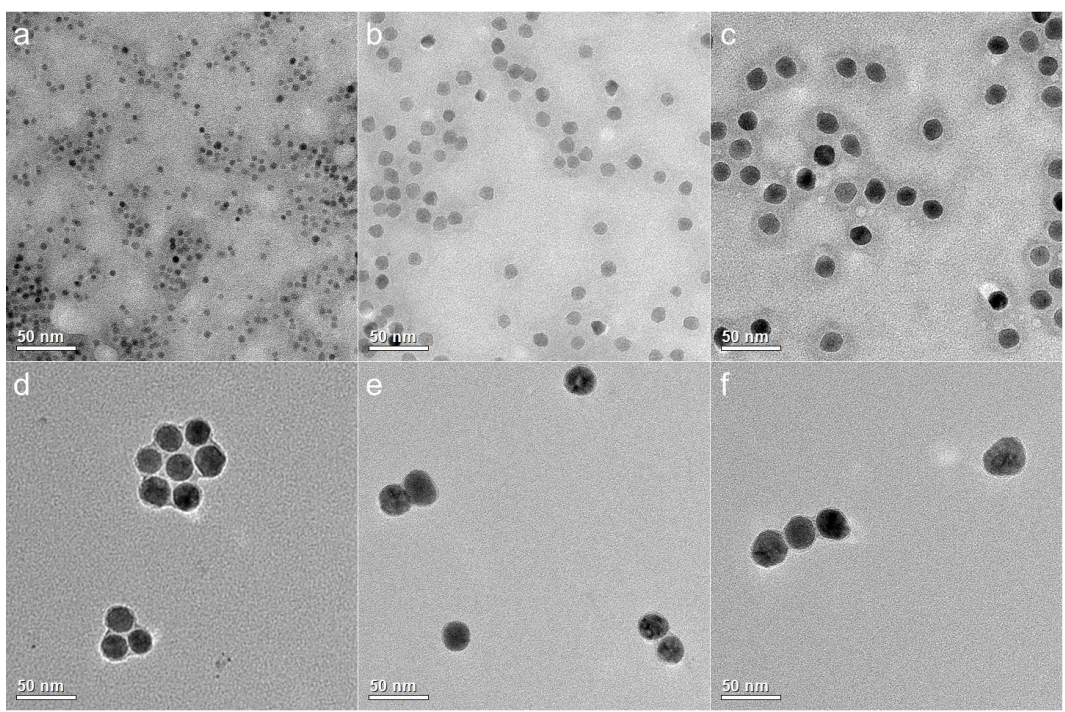

Figure S6-2. TEM images of SHA series MNPs. Scale bar represents 50 nm. (a) – (f) are SHA-5, SHA-10, SHA-15, SHA-20, SHA-25, and SHA-30, respectively.

## S7. Magnetic dynamic responses and higher harmonic models.

In the presence of alternating magnetic fields, MNPs are magnetized and their magnetic moments tend to align with the fields. For a monodispersed, non-interacting MNP system, the static magnetic response obeys the Langevin model:

$$M_D(t) = m_s c L(\xi),$$

where,

$$L(\xi) = \coth \xi - \frac{1}{\xi}, \xi = \frac{m_s H(t)}{k_B T}$$

The MNPs are characterized by magnetic core diameter $D$, saturation magnetization $M_s$ and particle concentration $c$, assuming MNPs are spherical without mutual interactions. The magnetic moment of each particle is expressed as $m_s = M_s V_c = M_s \pi D^3/6$, where $V_c$ is volume of the magnetic core, $\xi$ is the ratio of magnetic energy over thermal energy, $k_B$ is Boltzmann constant, and $T$ is the absolute temperature in Kelvin. $H(t)$ is the external magnetic driving fields represented in Figure S2.

However, artificially synthesized MNPs do not yield identical diameters. Another reasonable and commonly used approach is the log-normal size distribution model. Here, to simplify the model, we are assuming the MNPs are with identical core diameter of $D$.

Harmonics model of mono-frequency MPS system:

Taylor expansion of $M_D(t)$ shows the major harmonic components:

$$\frac{M_D(t)}{m_s c} = L\left(\frac{m_s H(t)}{k_B T}\right)$$

$$= \frac{1}{3}\left(\frac{m_s}{k_B T}\right) H(t) - \frac{1}{45}\left(\frac{m_s}{k_B T}\right)^3 H(t)^3 + \frac{2}{945}\left(\frac{m_s}{k_B T}\right)^5 H(t)^5 + \cdots$$

$$= \cdots + \left[\frac{1}{180} A^3 \left(\frac{m_s}{k_B T}\right)^3 - \frac{1}{1512} A^5 \left(\frac{m_s}{k_B T}\right)^5 + \cdots\right] \times \sin[2\pi \cdot 3f \cdot t]$$

$$+ \left[\frac{1}{7560} A^5 \left(\frac{m_s}{k_B T}\right)^5 + \cdots\right] \times \sin[2\pi \cdot 5f \cdot t]$$

$$+ \cdots$$

The higher odd harmonics are expressed as:

$$M_D(t)|_{3rd} \approx \frac{m_s c}{180} A^3 \left(\frac{m_s}{k_B T}\right)^3 \times \sin[2\pi \cdot 3f \cdot t]$$

$$M_D(t)|_{5th} \approx \frac{m_s c}{7560} A^5 \left(\frac{m_s}{k_B T}\right)^5 \times \sin[2\pi \cdot 5f \cdot t]$$

Harmonics model of dual-frequency MPS system:

Taylor expansion of $M_D(t)$ shows the major frequency mixing components:

$$\frac{M_D(t)}{m_s c} = L\left(\frac{m_s H(t)}{k_B T}\right)$$

$$= \frac{1}{3}\left(\frac{m_s}{k_B T}\right) H(t) - \frac{1}{45}\left(\frac{m_s}{k_B T}\right)^3 H(t)^3 + \frac{2}{945}\left(\frac{m_s}{k_B T}\right)^5 H(t)^5 + \cdots$$

$$= \cdots + \left[-\frac{1}{60} A_H A_L^2 \left(\frac{m_s}{k_B T}\right)^3 + \cdots\right] \times \cos[2\pi(f_H \pm 2f_L)t]$$

$$+ \left[\frac{1}{1512} A_H A_L^4 \left(\frac{m_s}{k_B T}\right)^5 + \cdots\right] \times \cos[2\pi(f_H \pm 4f_L)t]$$

$$+ \cdots$$

The mixing frequency components are found at odd harmonics exclusively:

$$M_D(t)|_{3rd} \approx \frac{-m_s c}{60} A_H A_L^2 \left(\frac{m_s}{k_B T}\right)^3 \times \cos[2\pi(f_H + 2f_L)t]$$

$$M_D(t)|_{5th} \approx \frac{m_s c}{1512} A_H A_L^4 \left(\frac{m_s}{k_B T}\right)^5 \times \cos[2\pi(f_H + 4f_L)t]$$

According to the Faraday's law of induction, the induced voltage in a pair of pick-up coils is expressed as:

$$u(t) = -S_0 V \frac{d}{dt} M_D(t)$$

where $V$ is volume of MNP suspension. Pick-up coil sensitivity $S_0$ equals to the external magnetic field strength divided by current.

The static magnetic response mode (the Langevin model) discussed above is unable to describe the dynamic responses of MNPs suspended in solution. Herein, Néel and Brownian relaxation models are introduced to complete the model.

## S8. Brownian and Néel relaxations.

Figure S8 shows that small MNPs relax via Néel process whereas larger MNPs relax via Brownian process. The cut off size is around 13 nm.

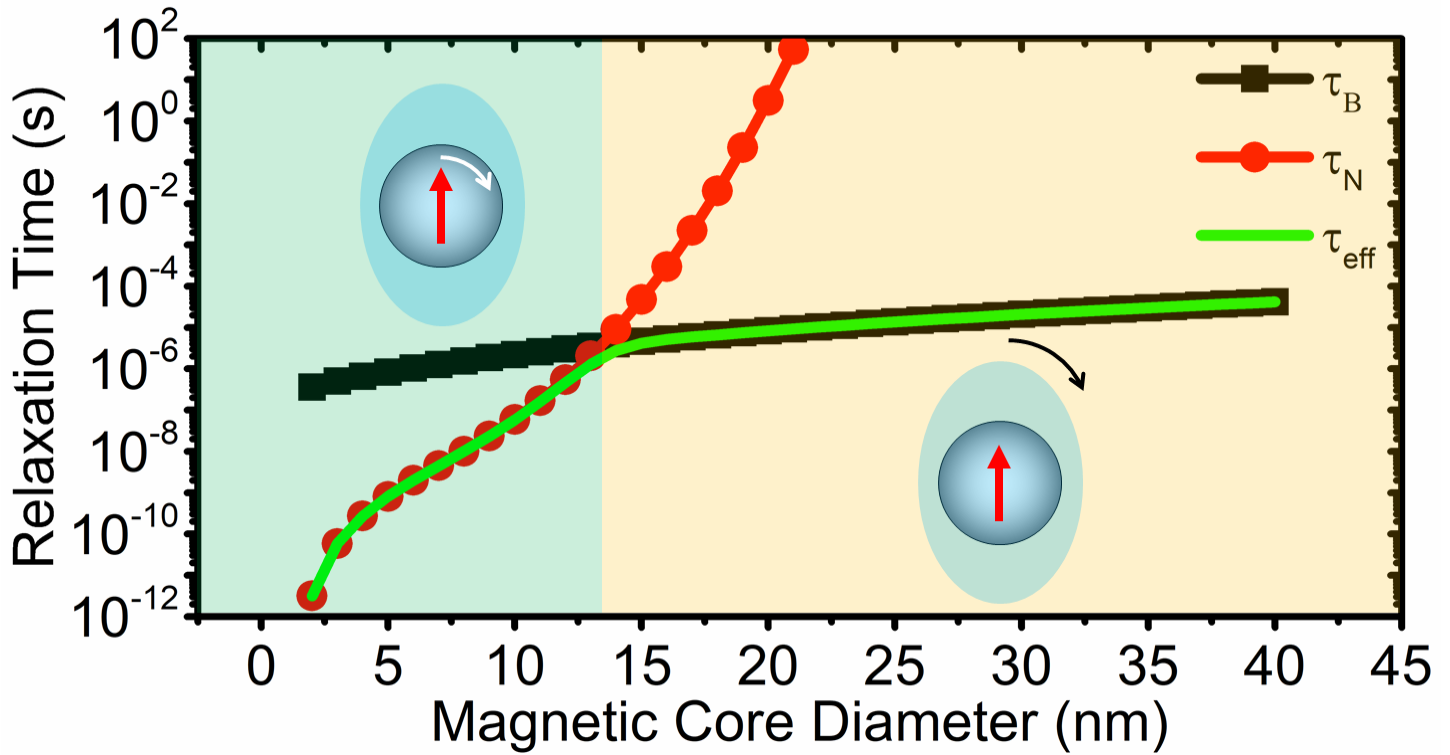

Figure S8. Simulated Brownian, Néel, and effective relaxation time as function of MNP core diameters. The crystal asymmetry on the surface of nanoparticles (also called "magnetically dead layer") yields a smaller saturation magnetization $M_s$ and a larger anisotropy constant $K_{eff}$ than the bulk materials [31]. Due to this surface spin-canting effect, $K_{eff}$ and $M_s$ are calculated for different sizes of MNPs, polymer coating layer thickness is assumed to be $d = 4\ nm$, solution viscosity is assumed to be $\eta = 1\ cp$.

**S9**. Phase angles of higher harmonics monitored under low frequency driving field for SHA series samples.

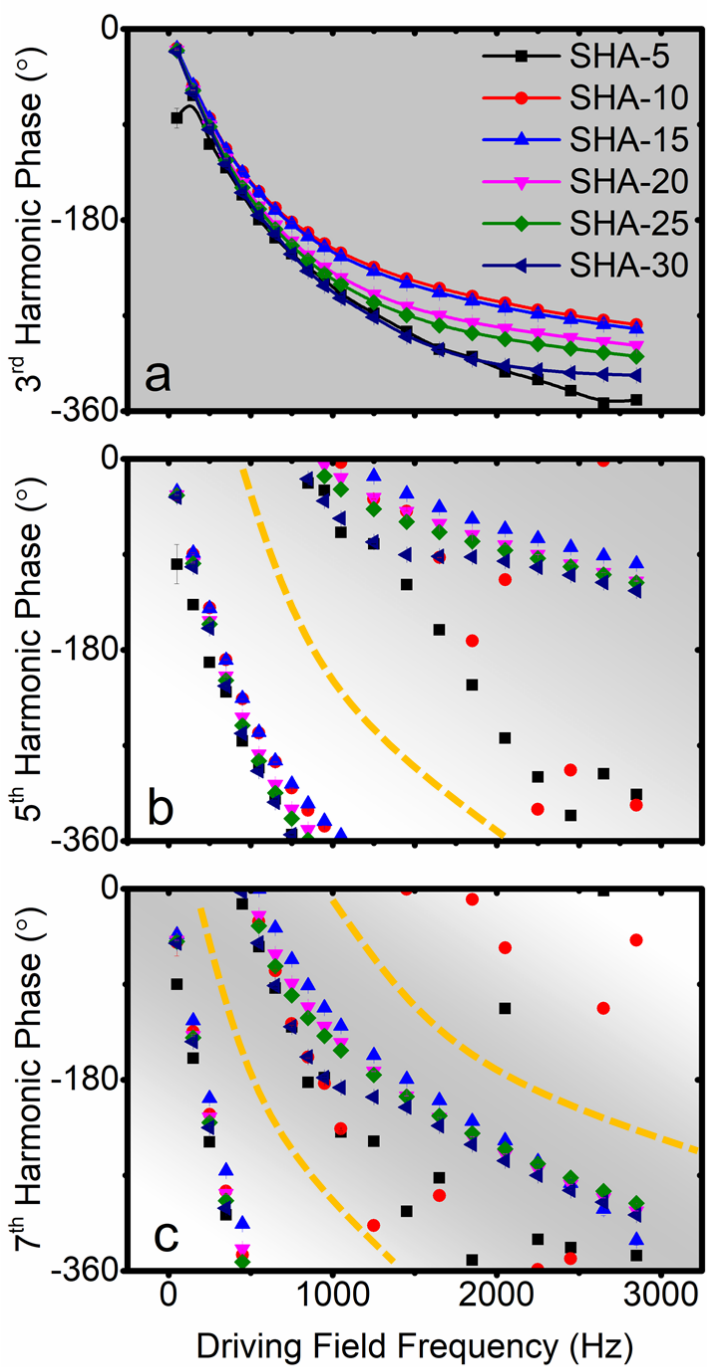

Figure S9. Phase angles of higher harmonics monitored under low frequency driving field for SHA series samples. (a), (b) and (c) summarizes the phase angles of the 3$^{rd}$, the 5$^{th}$, and the 7$^{th}$ harmonics under different driving field frequencies.

**S10.** Phase angles of higher harmonics monitored under high frequency driving field for SHA series samples.

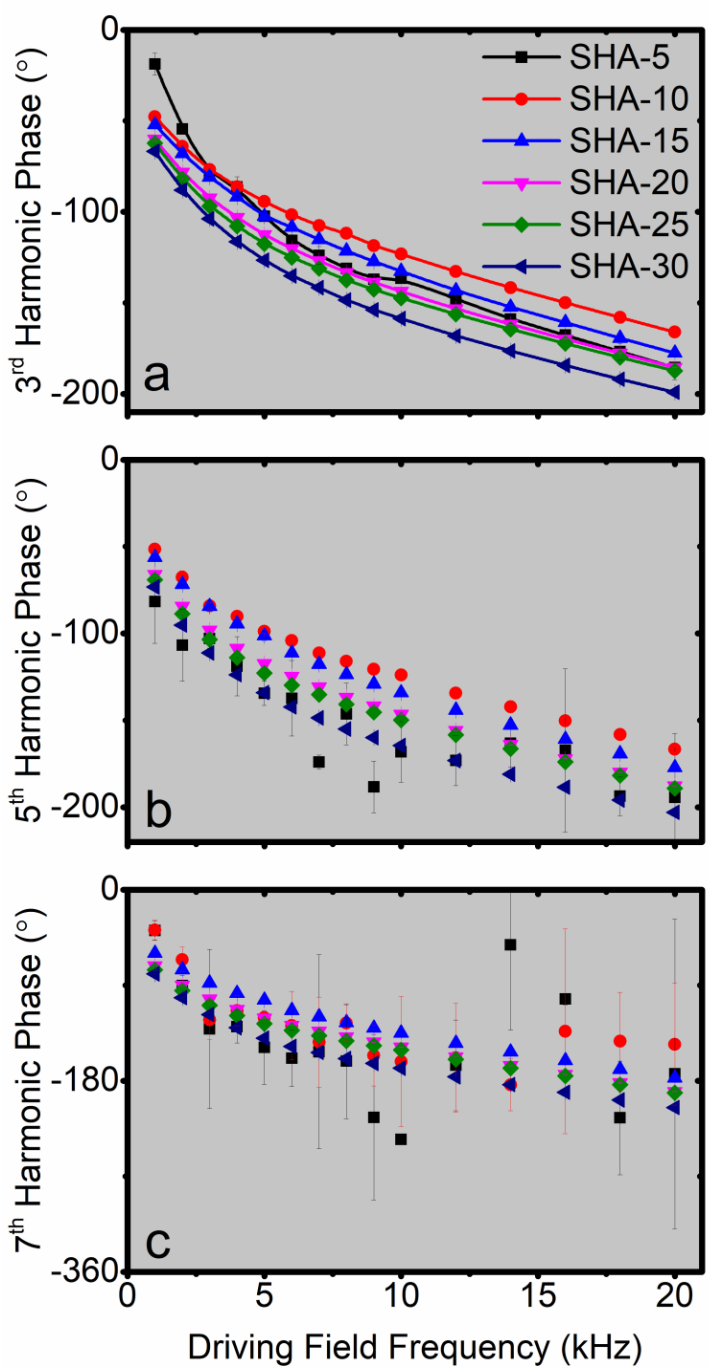

Figure S10. Phase angles of higher harmonics monitored under high frequency driving field for SHA series samples. (a), (b) and (c) summarizes the phase angles of the 3$^{rd}$, the 5$^{th}$, and the 7$^{th}$ harmonics under different driving field frequencies.